\documentclass[referee]{raa}            

\usepackage{graphicx,times}             
\usepackage{natbib}
\usepackage{amssymb,amsmath}
\bibpunct{(}{)}{;}{a}{}{,}

\usepackage[pagebackref=true]{hyperref}

\graphicspath{{./}{fig/}}

\begin{document}

    \title{
        Dynamical Origin of (469219) Kamo`oalewa of Tianwen-2 Mission from the Main-Belt:
        $\nu_6$ Secular Resonance, Flora Family or 3:1 Resonance with Jupiter
    }

    \volnopage{Vol.0 (202x) No.0, 000--000}      
    \setcounter{page}{1}          

    \author{Yandong Wang 
        \inst{1, 2}
    \and Shoucun Hu
        \inst{1, 2}
    \and Jianghui Ji
        \inst{1, 2}
    \and Jiajun Ying
        \inst{1, 2}
    }

    \institute{
        Purple Mountain Observatory,
        Chinese Academy of Sciences,
        Nanjing 210023, China; {\it jijh@pmo.ac.cn} \\
        \and
        School of Astronomy and Space Science,
        University of Science and Technology of China,
        Hefei 230026, China \\
        \vs\no
    {\small Received 202x month day; accepted 202x month day}
    }

    \abstract{
        China's \emph{Tianwen-2} mission, launched on 29 May 2025, targets the near-Earth object (469219) Kamo`oalewa, an Earth quasi-satellite trapped in a 1:1 mean-motion resonance
        with our planet. Determining the origin of Kamo`oalewa is central to understanding the formation pathways and dynamical evolution of Earth's quasi-satellite population.
        Here we show a strong possibility of main-belt origin for Kamo`oalewa using long-term
        dynamical simulations. We examine three candidate source regions: the $\nu_6$ secular resonance ($\nu_6$), the 3:1 mean-motion resonance with Jupiter (3:1J MMR), and the Flora family.
        A total of 42\,825 test particles were integrated over 100 Myr. We find that asteroids from
        all three regions can be transported onto Kamo`oalewa-like orbits, albeit with markedly
        different efficiencies. Particles originating near the $\nu_6$ show the highest transfer probability (3.31\%), followed by the Flora family (2.54\%) and the 3:1J MMR (0.39\%). We further identify representative dynamical pathways linking these source
        regions to Earth quasi-satellite orbits. The \emph{Tianwen-2} spacecraft is expected to rendezvous with Kamo`oalewa in 2026, performing close-proximity operations and returning
        samples. The mission will provide decisive observational constraints on the asteroid's composition and physical properties, offering a critical test of its proposed origin.
    }

   \authorrunning{}            
   \titlerunning{Dynamical Origin of (469219) Kamo`oalewa from Main-Belt}  

   \maketitle

\section{Introduction}           
\label{sect:intro}

    China's \emph{Tianwen-2} mission is designed to advance the exploration of small bodies by
    conducting rendezvous and sample-return operations at a near-Earth asteroid.
    Launched on 2025 May 29, the spacecraft targets (469219) Kamo`oalewa (provisional
    designation $\text{2016~HO}_3$), discovered on 2016 April 27 by the Pan-STARRS1 survey
    telescope at the Haleakala Observatory in Hawaii, which is known as an Earth quasi-satellite
    temporarily trapped in a 1:1 mean-motion resonance with Earth \citep{Marcos2016}.
    Such objects provide a natural laboratory for studying the dynamical coupling between the
    near-Earth population and the main-belt. Through close-proximity observations and the
    return of pristine material, \emph{Tianwen-2}  will place new constraints on the physical
    properties and origin of quasi-satellites.

    In the restricted three-body problem, quasi-satellites
    (also referred to as retrograde satellites; \citealt{Jackson1913})
    constitute a special class of small bodies locked in 1:1 mean-motion resonance with their host
    planet, with the relative mean longitude librating around $0^\circ$\citep{Marcos2016}.
    In a heliocentric frame of reference that rotates with host planet,
    these objects exhibit satellite-like apparent motion, orbiting the planet;
    however, their orbits always remain outside the radius of the Hill sphere of the host planet
    and they are not gravitationally bound. Dynamically, they therefore retain heliocentric
    orbits rather than true satellite states \citep{Mikkola2006}.
    As of November 2025, only 8 Earth quasi-satellites have been confirmed, among which (469219)
    Kamo`oalewa is one of the most dynamically stable
    Earth co-orbitals currently known \citep{Marcos2016, Marcos2025}. Its distinctive physical and
    dynamical properties make it an important target for studies of Earth co-orbitals.

    Kamo`oalewa has an absolute magnitude $H=24.33$~mag and a semimajor axis $a = 1.00098$~au
    {(at epoch JD~2461000.5)}, yielding an orbital period very close to that of Earth.
    Its orbital eccentricity and inclination are $e = 0.10237$ and $i = 7.80322^\circ$,
    respectively. Broadband color photometry and visible spectra indicate that its taxonomic class
    is consistent with S-type silicate asteroids, but with a redder spectral slope than
    typical S-type objects \citep{Sharkey2021}, suggesting that it may have experienced
    substantial space weathering \citep{Zhang2024}.
    Assuming a geometric albedo of 0.1, \citet{Zhang2024} estimated an
    equivalent diameter of approximately 57 m. In addition, Kamo`oalewa exhibits an exceptionally
    rapid rotation period of about 28 min, well below the 2.2-h spin barrier for cohesionless
    small bodies \citep{Pravec2000, chang2017}. This implies that it may possess a relatively
    coherent monolithic interior, possibly mantled by a regolith layer composed of millimeter- to
    centimeter-sized particles \citep{li_shape_2021, cheng2024, Liu2024, Ren2024, YING2025}.

    From a dynamical perspective, numerical simulations by \citet{Marcos2016} showed that
    Kamo`oalewa undergoes recurrent and stable transitions between the quasi-satellite and
    horseshoe configurations on Myr timescales. Its current quasi-satellite phase is estimated to
    have begun about 100 yr ago and is expected to end in roughly 300 yr.
    Furthermore, \citet{Fenucci2021} explored the long-term dynamical consequences of
    Yarkovsky-induced semimajor-axis drift for Kamo`oalewa under different surface
    thermal-property scenarios. They showed that while the Yarkovsky effect has little influence
    on the short-term dynamics of the asteroid, it can significantly affect its long-term stability
    for low surface thermal conductivity, accelerating its escape from the Earth co-orbital region.
    Building on this theoretical framework, \citet{Liu2022} reported the first observational
    detection of the Yarkovsky effect acting on Kamo`oalewa based on ground-based optical
    astrometry. Their results confirmed that the effect is negligible on short timescales but
    reduces the asteroid's residence time in the Earth co-orbital state, in agreement with earlier
    numerical predictions. Further advances were made by \citet{Hu2023}, who derived new
    estimates of the non-gravitational parameter $A_2$ and explored the propagation of orbital
    instability, suggesting that the quasi-satellite configuration itself may play a key role in
    limiting orbital uncertainty. Using updated astrometric observations, \citet{Fenucci2025}
    further constrained the non-gravitational parameter $A_2$ and the thermal inertia of
    Kamo`oalewa, thereby improving the predicted targeting accuracy of the \emph{Tianwen-2} spacecraft
    during its encounter.

    Unlike long-lived co-orbital populations such as the Jovian Trojans, the complex multi-body
    perturbations in the inner Solar System lead to a much higher degree of orbital chaos for
    NEOs \citep{Morais2002}. As a result, co-orbital small bodies associated
    with terrestrial planets exhibit rich and distinctive dynamical behaviors, and their origins
    and evolutionary pathways provide important insights onto the dynamical structure of the
    inner Solar System.

    Recently, the origin of Kamo`oalewa has been the subject of several dedicated studies.
    Based on analysis of visible and near-infrared reflectance spectral, \citet{Sharkey2021} found
    that the spectral characteristics of this object, particularly the position and shape of the
    $1\,\mu\mathrm{m}$ absorption band, are highly similar to those of lunar regolith samples
    returned by the Apollo mission, and for the first time proposed a lunar-origin hypothesis for
    Kamo`oalewa.

    Using $N$-body numerical simulations, \citet{Castro2023} examined the dynamical feasibility of
    lunar impact ejecta being injected into the Earth co-orbital region. They showed that lunar
    ejecta released from the trailing side of the Moon at velocities slightly exceeding the lunar
    escape speed have a non-negligible probability of being captured as Earth co-orbital objects,
    and may even evolve into quasi-satellite orbits.
    Combining spectral constraints with dynamical analyses, \citet{Jiao2024} argued that
    Kamo`oalewa most likely originated from the young lunar impact crater Giordano Bruno.
    In contrast, \citet{Zhu2025} based on more detailed spectral comparisons and impact-dynamical
    simulations, suggested that the Tycho crater may represent a source region
    more consistent with both observational and dynamical constraints.

    Although previous studies have provided evidence supporting a lunar origin for Kamo`oalewa
    from both spectroscopic observations and $N$-body numerical simulations, this scenario remains
    debated. Based on spectral analyses, \citet{Fenucci2021} were among the first to suggest that
    Kamo`oalewa may be related to Barbarian asteroids in the main-belt and could have migrated to
    its current orbit through the 3:1J~MMR{, located at approximately 2.5~au, which
    corresponds to the commensurability $n/n_J = 3$, where $n$ and $n_J$ are the mean motions of
    the asteroid and Jupiter, respectively \citep{Wisdom1985, Gladman1997, Bottke2002, Granvik2018}}.
    Using spectral comparisons combined with orbital dynamical calculational method,
    \citet{Zhang2024} argued that Kamo`oalewa is compositionally more consistent with LL
    chondrites and has a $72 \pm 5\%$ probability of originating from the $\nu_6$ region,
    {where the precession frequency of the asteroid's longitude of perihelion $g$ is equal to mean
    precession frequency of Saturn's longitude of perihelion $g_6$ \citep{Froeschle1989, Morbidelli2002}}.
    The main-belt, particularly its strong resonant regions, is widely regarded as the
    primary source of NEOs \citep{Binzel1992, Morbidelli2002, Morbidelli2003, Michel2005}.
    In this context, Earth co-orbital objects do not constitute an isolated dynamical population
    but may be connected with near-Earth and main-belt asteroids.
    It is therefore necessary to assess the dynamical feasibility of main-belt regions
    producing Kamo`oalewa-like orbital configurations.

    In addition, existing dynamical simulations investigating a lunar origin for Kamo`oalewa have
    considered only planetary gravitational perturbations and have not incorporated the
    Yarkovsky effect. For small bodies with diameters of only several tens of
    meters, this effect can significantly modify orbital evolution on Myr timescales,
    thereby altering the probability of entering and maintaining co-orbital configurations
    \citep{Fenucci2021, Liu2022}. These limitations leave substantial room for further
    investigation of the origin and dynamical evolution of Kamo`oalewa from a dynamical perspective.

    Motivated by this, we use numerical integrations that include Yarkovsky-driven forces to
    assess whether Kamo`oalewa could originate from the main-belt and to identify the
    dynamical pathways by which particles evolve onto Kamo`oalewa-like orbits.
    The \emph{Tianwen-2} spacecraft is scheduled to rendezvous with Kamo`oalewa in July 2026,
    marking China's first asteroid sample-return mission, and our results provide essential
    dynamical context for interpreting the mission's forthcoming observations.

    This paper is organized as follow: Section~\ref{sect:DandM} introduces the data and the numerical
    setup in this study. Section~\ref{sect:results} analysis the results and presents
    several representative evolutionary paths for our candidate sources.
    The conclusion and discussions are presented in Section~\ref{sect:conclusion}.

\section{Data and method}
\label{sect:DandM}

    \citet{Marcos2016} reported that the Lyapunov time of Kamo`oalewa is approximately 7500 yr,
    indicating a high degree of orbital chaos on relatively short timescales.
    Moreover, \citep{Morbidelli2020} pointed out that long-term backward orbital integrations
    cannot reliably trace the origin of small bodies. In light of these considerations,
    we adopt a forward orbital integrations approach combined with statistical analysis to assess the
    dynamical feasibility of objects from different main-belt sources evolving toward the
    current orbit of Kamo`oalewa.
    Our focus is on the statistical behavior of test particles, under the assumption that,
    for sufficiently large samples, the collective evolution of a large number of test particles
    can provide a representative picture of the likely dynamical pathways leading to Kamo`oalewa.

    Based on the debiased NEOs model proposed by \citet{Nesvorny2024},
    we estimated the potential dynamical sources of Kamo`oalewa.
    The results indicate that the $\nu_6$ and 3:1J MMR \citep{Ji2007, Nesvorny2023} in the
    main-belt are the most probable
    contributors, with corresponding probabilities of 71.7\% and 17.5\%, respectively.
    In addition, \citet{Zhang2024} further suggested that the Flora family also warrants consideration.
    Flora family corresponds to a major asteroid family in the inner main-belt located near
    $\nu_6$ and is considered a primary source of LL chondrite-like NEOs as well as Earth--Moon
    impactors \citep{Vokrouhlicky2017, Zhang2025}.
    Considering the above factors, we select three candidate sources in the main-belt,
    namely the $\nu_6$, the 3:1J MMR, and the Flora family, for further numerical simulations to
    investigate the dynamical origin of such orbits.

    The approximate boundaries of $\nu_6$ and 3:1J MMR are defined following the method of
    \citet{Nesvorny2023}, with Mars-crossing objects excluded.
    In view of the phase-space structure of the $\nu_6$ in the vicinity of the
    3:1J MMR, we restrict the upper inclination limit of $\nu_6$ to
    $17^\circ$. Members of Flora family are obtained from the NASA Planetary Data System
    \footnote{https://sbn.psi.edu/pds/resource/nesvornyfam.html} (see also \citealp{Nesvorny2020}).

    We select real asteroids located in the three candidate sources from the MPCORB.DAT
    catalog provided by the Minor Planet Center (version dated 2025 June 04) and use them as
    simulated particles. The initial orbital elements of all simulated particles are taken
    directly from this catalog, with an initial epoch of JD~2460800.5.

    It should be emphasized that our numerical simulations are not intended to investigate the
    detailed evolution of individual real asteroids. Instead, the observed asteroids are treated
    as samples of the corresponding sources, allowing us to extract statistically
    representative dynamical behaviors and underlying mechanisms in their long-term evolution.

\begin{figure}
   \centering
   \includegraphics[width=0.9\textwidth]{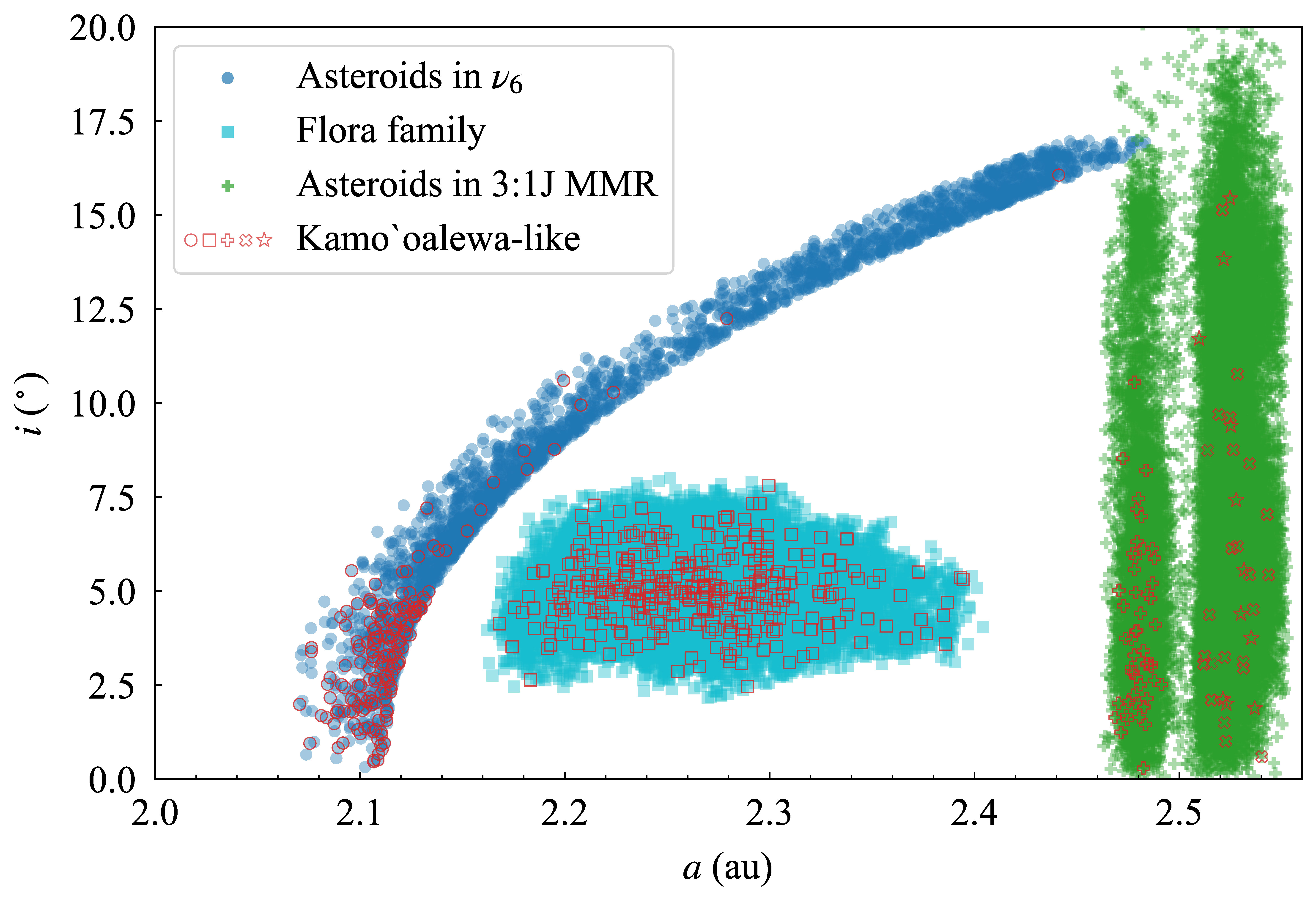}
   \caption{
        Initial distribution of simulated particles in the $a$--$i$ plane.
        Blue circles, cyan squares, and green plus symbols denote particles initially located near
        the $\nu_6$, the Flora family, and the \mbox{3:1J MMR}, respectively.
        Red open symbols indicate particles from the candidate sources that eventually evolve
        onto Kamo`oalewa-like orbits during the integrations.
        For particles initially associated with the 3:1J MMR, red open plus and cross symbols represent
        those evolving onto Kamo`oalewa-like orbits via the $\nu_6$ and \mbox{3:1J MMR}, respectively,
        while red star symbols denote particles that first cross the 3:1J MMR and subsequently
        evolve via the $\nu_6$.}
   \label{Fig1}
   \end{figure}

    The total number of sampled particles is 42\,825, including 7\,122 in the $\nu_6$, 13\,786 in
    the Flora family, and 21\,917 in the 3:1J MMR. The initial distribution of these particles in
    the $a$--$i$ plane is shown in Fig.~\ref{Fig1}.

    We perform numerical integrations of the simulated particles using the Hybrid integrator of
    the \texttt{MERCURY6} package \citep{Chambers1999}. Gravitational perturbations from all big
    bodies from Mercury to Pluto are included, while the Earth--Moon system is represented by its
    barycenter. The initial states of all perturbing bodies are taken from the JPL Ephemerides DE441
    \citep{Park2021} and are set at the same epoch as the simulated particles.

    The codes are modified to incorporate the Yarkovsky effect in this work. Although the acceleration induced by
    the Yarkovsky effect is much smaller than that due to planetary gravitational perturbations,
    it can lead to long-term variations in the semimajor axis $a$ by increasing or decreasing the
    orbital energy of small bodies on timescales of Myr to Gyr \citep{Bottke2006, vokrouhlicky2015}.
    A precise treatment of the Yarkovsky effect requires detailed knowledge of the physical
    properties of individual objects; however, given the limited constraints on the physical
    properties of Kamo`oalewa, we adopt a simplified Yarkovsky acceleration model following
    \citet{Deo2017} and \citet{Hu2023}:
    \begin{equation}\label{eq1}
        \boldsymbol{F}_\text{Yark} = A_2 \left( r_0 / r \right)^d \hat{\boldsymbol{t}}
    \end{equation}
    where $r_0 = 1$~au is a reference heliocentric distance, $r$ is the instantaneous heliocentric
    distance of the particle expressed in au, and $d$ is a parameter related to the thermophysical properties of
    the asteroid. In this work, we adopt $d=2$. The unit vector $\hat{\boldsymbol{t}}$ points in
    the transverse direction of the orbit, and $A_2$ is the non-gravitational transverse parameter,
    taken from the value reported by the JPL Horizons system for Kamo`oalewa,
    $A_2 = -1.3455 \times 10^{-13}\ \mathrm{au\ day^{-2}}$, corresponding to a retrograde
    rotation state, and is held constant throughout the integrations.
    {
        We adopt a single, fixed value of $A_2$, applied uniformly to all test particles
        in our simulations for Kamo`oalewa.
    }

    During the integrations, all simulated particles are treated as massless bodies and are
    subject only to gravitational perturbations from the massive bodies and the Yarkovsky effect.
    A simulated particle is removed from the integrations if it collides with any massive body or
    if its heliocentric distance exceeds 100~au.

    Following the approach of \citet{Bottke2015}, we apply a modified criterion to identify
    particles that evolve onto Kamo`oalewa-like orbits. Specifically, we compute the mean orbital
    elements $(\overline{a},\overline{e},\overline{i})$ of Kamo`oalewa over a time interval of
    $2\times10^4$~yr, yielding (0.99989~au, 0.09934, $7.57547^\circ$).
    This criterion is designed to reduce the influence of short-term fluctuations in the
    osculating orbital elements on the classification.
    A simulated particle is classified as Kamo`oalewa-like if its evolving orbital elements satisfy
    $\Delta a \leq 0.01$~au, $\Delta e \leq 0.1$, and $\Delta i \leq 5^\circ$,
    where $\Delta a$, $\Delta e$, and $\Delta i$ denote the absolute differences
    from the reference mean values. {Although the adopted criteria are broader than those
    used in previous studies (e.g., \citealt{Bottke2015, Zhou2019}), given that  Kamo`oalewa
    undergoes transitions among different co-orbital states, its orbital elements can vary over
    a relatively broad range with time \citep{Marcos2016}. Accordingly, we adopt a relatively wide set of
    criteria to encompass the temporal variability of its orbital elements over a finite interval,
    rather than to match only its instantaneous present-day orbit. Our aim is not to reproduce
    the exact current quasi-satellite configuration of Kamo`oalewa, but to identify simulated
    particles whose orbital evolution is comparable to that of Kamo`oalewa over a certain timescale.}

\section{results}
\label{sect:results}
    After including the Yarkovsky effect, we integrated a total of 42\,825 real asteroids selected
    from three candidate main-belt sources: the $\nu_6$, the 3:1J MMR, and the Flora family for a
    duration of 100~Myr. We then identified particles that escape from these regions and
    subsequently evolve onto Kamo`oalewa-like orbits.

    In total, 672 simulated particles from the three candidate sources are found to reach
    Kamo`oalewa-like orbits during the integrations. Among them, 236 originate from the $\nu_6$,
    350 from the Flora family, and 86 from the 3:1J MMR, corresponding to fractions of 3.31\%,
    2.54\% and 0.39\% of their respective source populations. These results indicate that,
    although \mbox{Kamo`oalewa-like} orbits occupy only a limited region in the present-day phase space,
    particles originating from the main-belt can still reach such configurations.

    The distribution of Kamo`oalewa-like particles in initial orbital-parameter space shows additional
    features. Among particles originating from the $\nu_6$, about 96.2\% are found
    at inclinations lower than $7.8^\circ$ (the present inclination of Kamo`oalewa),  whereas the
    corresponding fraction for the 3:1J MMR is about 83.7\%.
    This distribution indicates that, for the two dynamical source regions considered here,
    only particles with specific initial orbital parameters in phase space{---particularly those
    with lower initial inclinations---}are able to evolve onto Kamo`oalewa-like orbits.

    \begin{figure}
    \centering
    \includegraphics[width=1\textwidth]{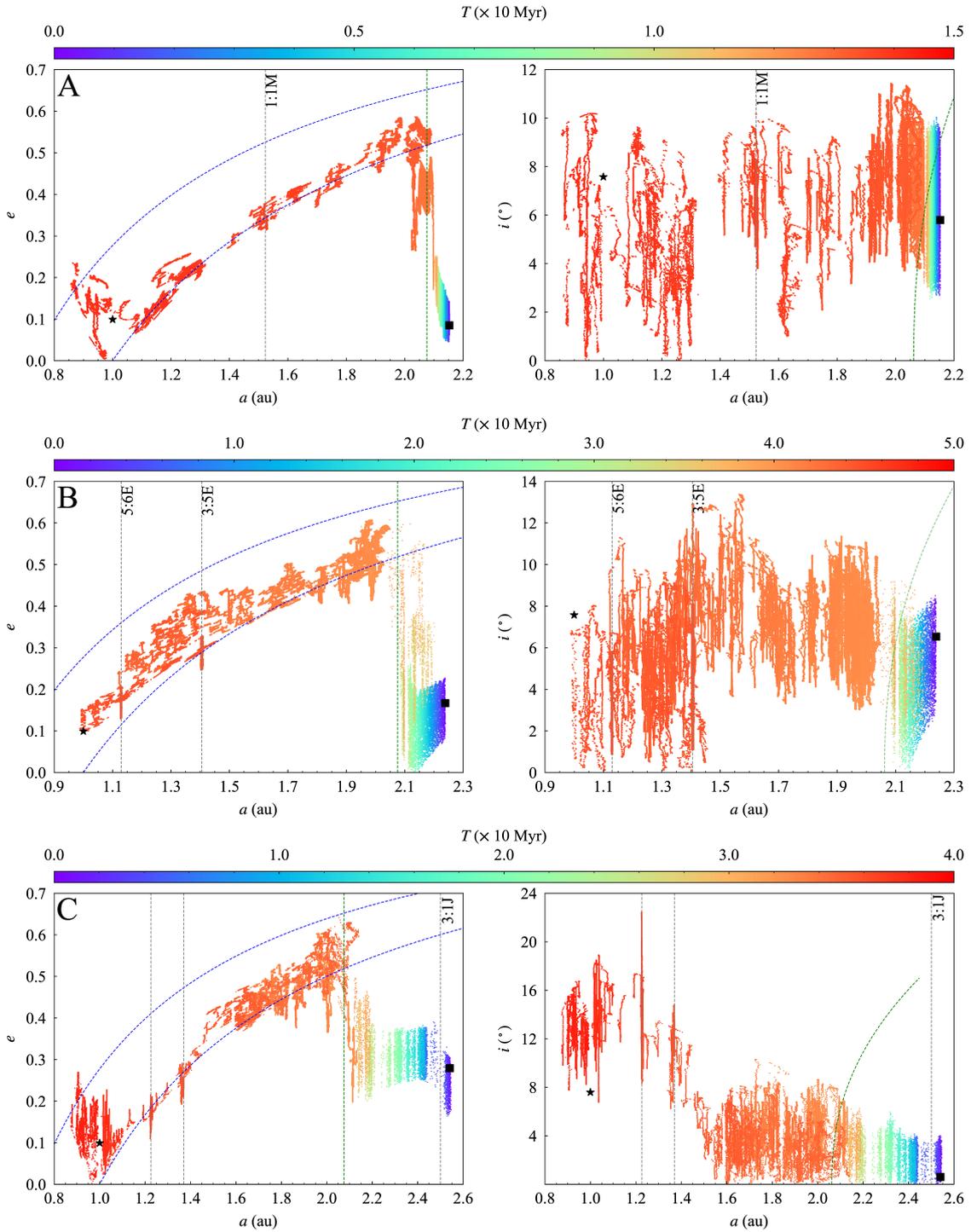}
    \caption{
        From top to bottom, the A, B and C panels show the representative orbital evolutions of
        simulated particles from the $\nu_6$, the Flora family, and the 3:1J MMR, respectively,
        from the main-belt (time $T=0$) to Kamo`oalewa-like orbits in the $a$--$e$ (left)
        and $a$--$i$ (right) planes.
        Colors indicate time progression from purple (early stages) to red
        (approaching Kamo`oalewa-like orbits).
        Black squares mark initial particle positions, and the black stars denote the current
        position of Kamo`oalewa.
        Gray dashed lines indicate representative mean-motion resonances, with unlabeled ones
        corresponding to closely spaced resonances.
        Green and blue dashed lines show the approximate location of the $\nu_6$ and the loci
        where the perihelion distance equals that of the Earth or Venus, respectively.
        For clarity, the scattered points are shown after down-sampling.
    }
    \label{Fig2}
    \end{figure}

    During their dynamical evolution, simulated particles exhibit relatively regular orbital
    behavior while residing in the main-belt, but become increasingly chaotic after entering the
    near-Earth region (Fig.~\ref{Fig2}). Overall, their evolution is governed by a combination of
    gravitational scattering by the massive bodies and episodes of temporary resonance capture,
    primarily involving two-body and three-body mean-motion resonances with Earth, Mars, and
    Jupiter. This general evolutionary behavior is consistent with the results reported by
    \citet{Gladman1997} and \citet{Michel2005}.

    Fig.~\ref{Fig2} illustrates the representative evolutionary behaviors of particles originating
    from the three candidate sources. In general, particles spend most of their
    evolutionary lifetimes in the main-belt during their migration toward the near-Earth region.
    Once they approach the main-belt escape channels---dominated in our simulations by the
    $\nu_6$ and the 3:1J MMR---they reach the near-Earth region on timescales of
    $10^{5}$--$10^{6}$~yr.

    Panel~A of Fig.~\ref{Fig2} shows a representative evolutionary pathway of a particle originating near
    the $\nu_6$. The particle initially resides at $a=2.15$~au and $i=5.8^\circ$
    and is locked in the $\nu_6$ with the resonant angle
    $\Delta \varpi = \varpi - \varpi_6$ oscillating around $180^\circ$, where $\varpi$ and
    $\varpi_6$ denote the longitudes of perihelion of the particle and Saturn, respectively.
    This configuration corresponds to a relatively stable anti-aligned state with Saturn
    \citep{Carruba2022}, allowing the particle to remain in the $\nu_6$ for
    approximately 12~Myr. During this stage, the eccentricity exhibits oscillatory behavior with
    a slow overall increase, remaining insufficient to reduce the perihelion distance to
    Earth-crossing values. Over the subsequent $\sim$1~Myr, the eccentricity rapidly increases
    to $\sim0.6$, driving the particle into an Earth-crossing orbit and eventually causing it to
    escape from the resonance following a close encounter with Earth.

    After leaving the resonance, the particle evolves while its perihelion remains near $q\simeq1$.
    During this phase, it experiences multiple episodes of temporary resonance capture.
    As shown in the right panel of Fig.~\ref{Fig2}A, these resonances modulate not only the
    eccentricity but also the inclination, leading to oscillations in $i$ while the $a$
    remains nearly constant. The capture durations range from a few thousand to several tens of
    thousands of years. The longest capture occurs near $a \simeq 1.524$~au, corresponding to a
    1:1 mean-motion resonance with Mars (marked by a vertical dashed line), where the particle
    remains for nearly $3\times10^{4}$~yr and its inclination decreases from $\sim8^\circ$ to  $\sim4^\circ$.

    Because a negative value of the Yarkovsky parameter $A_2$ is adopted in the simulations,
    the semimajor axis of the particles exhibits a long-term inward drift \citep{Farnocchia2013}.
    As a result, particles initially located in the Flora family experience a gradual decrease
    in semimajor axis, causing their orbits to drift in $a$ toward the characteristic semimajor
    axis of the $\nu_6$, as illustrated in Fig.~\ref{Fig2}B.
    After entering the $\nu_6$ at $\sim$30~Myr, the eccentricity is excited to
    $\sim$0.4 within 3 Myr, allowing the particle to become Mars-crosser. Subsequent close
    encounters with Mars further drive the particle into an Earth-crossing orbit. The later
    evolution of this particle closely resembles that of particles initially located near the
    $\nu_6$. In Fig.~\ref{Fig2}B, vertical dashed lines mark several episodes of temporary
    resonance capture near $a\simeq1.225$~au and $a\simeq1.37$~au, where dense resonant structures
    delay the delivery of particle onto Kamo`oalewa-like orbits.

     The asteroids initially located near the 3:1J MMR exhibit more diverse behaviors owing to their
     distribution on both sides of the resonance (Fig.~\ref{Fig1}). Particles starting on the
    right-hand side of the resonance (marked by red open crosses) experience inward semimajor-axis
    drift driven by the Yarkovsky effect, {as only a negative value of $A_2$ is adopted in our
    simulations. Given the strong efficiency of the 3:1J~MMR in exciting eccentricities, particles
    entering the resonance have a much higher probability of reaching planet-crossing eccentricities
    than those remaining outside it. In our simulations, particles drifting into the 3:1J~MMR from
    the right-hand side of the resonance typically experience rapid eccentricity growth, reaching
    values of $e \gtrsim 0.6$ within less than 1~Myr, thereby becoming Earth-crossers, consistent
    with the results of \citet{Morbidelli1998}.} In contrast, particles initially located on the
    left-hand side of the resonance (marked by red open plus symbols) evolve in a manner similar to
    those from the Flora family, but require substantially longer migration times typically at
    least 50~Myr in our simulations to reach the $\nu_6$.

    We further identify a third evolutionary channel, in which some particles initially located
    at $a>2.5$~au (marked by red open stars) are not stably captured by the 3:1J MMR but instead cross
    it rapidly within $\sim$0.1~Myr. During the crossing, their eccentricities undergo modest
    oscillations with amplitudes of $\sim$0.2. In our simulations, 11 particles reaching
    Kamo`oalewa-like orbits follow this pathway. This behavior indicates that resonance capture at
    the 3:1J MMR is not inevitable for particles approaching from the right-hand side, but instead
    depends on factors such as the migration rate and/or initial orbital configuration.
    Similar resonance-crossing processes have been invoked to explain the origin of V-type
    asteroids located outside the Vesta family by \citet{Folonier2014}, suggesting a common
    underlying dynamical mechanism. The subsequent evolution of particles that cross the 3:1J MMR
    is similar to that of particles initially located in the Flora family, as well as particles
    on the left-hand side of this resonance, with representative examples shown in
    Fig.~\ref{Fig2}C. {All these particles follow similar evolutionary pathways, reaching
    Kamo`oalewa-like orbits via the $\nu_6$, as evidenced by the libration of $\Delta \varpi$.}

    \begin{figure}
    \centering
    \includegraphics[width=0.7\textwidth]{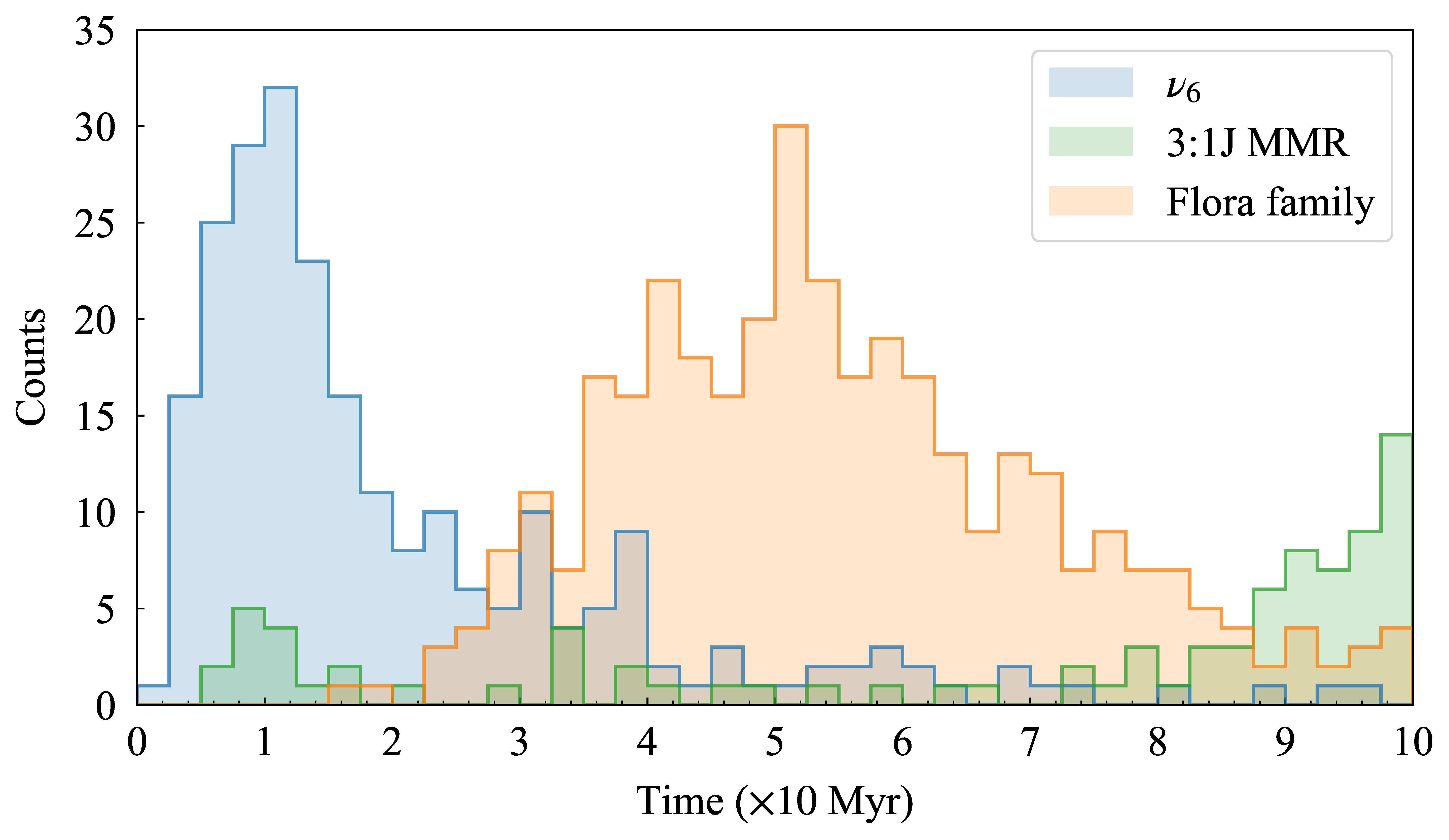}
    \caption{
        Statistical distribution of the time required for simulated particles originating from the
        three candidate source regions to evolve from their initial locations into
        Kamo`oalewa-like orbits.
    }
    \label{Fig3}
    \end{figure}

    Fig.~\ref{Fig3} summarizes the time required for simulated particles originating from the
    three candidate sources to evolve onto Kamo`oalewa-like orbits. The results show that
    during the early stage of the integrations (the first $\sim$30~Myr), Kamo`oalewa-like objects
    are predominantly supplied by particles initially located near the $\nu_6$,
    indicating that this region is more efficient at delivering particles onto Kamo`oalewa-like
    orbits at early times.

    As the integrations proceed, the relative contribution from the Flora family gradually
    increases and eventually becomes dominant. This transition can be attributed to the long-term
    action of the Yarkovsky effect, which continuously drives particles from the Flora family inward
    in semimajor axis, allowing them to progressively enter the $\nu_6$ and thereby substantially
    enhancing their probability of evolving onto Kamo`oalewa-like orbits.

    In contrast, the time distribution near the 3:1J MMR exhibits a
    distinct pattern. As discussed above, particles contributing at shorter timescales mainly
    originate from the right-hand side of the 3:1J MMR and evolve onto Kamo`oalewa-like orbits
    after their eccentricities are directly excited during resonance interaction.
    In our simulations, about 50\% of particles transported via the 3:1J MMR reach Kamo`oalewa-like
    orbits within 20~Myr. The obvious increase appearing after 80~Myr in Fig.~\ref{Fig3} is produced by a
    combination of particles initially located on the left-hand side of the resonance and
    particles originating on the right-hand side that undergo resonance crossing
    before evolving further.

    {Recent spectral analyses by \citet{Zhang2024} suggest that Kamo`oalewa is a space-weathering-matured
    object, with a relevant timescale of at least
    $\sim$50~Myr. In combination with our results shown in Fig.~\ref{Fig3}, Kamo`oalewa-like objects show a
    preference for originating from the Flora family or the 3:1J MMR. Nevertheless, a contribution from the
    $\nu_6$ source region cannot be completely ruled out.}

    Taken together, our results indicate that, given the established transport of main-belt
    objects into near-Earth space, main-belt regions can provide dynamically viable
    pathways leading to Kamo`oalewa-like orbits. Although such evolutionary outcomes
    represent only a small fraction of the overall population, their existence demonstrates that
    the main-belt origin for Kamo`oalewa-like objects is highly likely on dynamical grounds.

\section{Conclusion and discussions}
\label{sect:conclusion}
    Previous studies on the origin of (469219) Kamo`oalewa have largely focused on a lunar
    impact-ejecta scenario, with significant insights gained from spectroscopic and
    dynamical analyses.
    Yet it remains unclear whether known near-Earth object reservoirs, particularly the main-belt,
    can dynamically supply Kamo`oalewa-like objects to the near-Earth region.
    Here, we explore this possibility and investigate the main-belt pathways that could deliver  such bodies.
    Incorporating the Yarkovsky effect, we carry out long-term numerical integrations with the
    \texttt{MERCURY6} package for 42\,825 asteroids drawn from three main-belt source regions:
    the $\nu_6$, the 3:1J MMR, and the Flora family.
    Our goal is to systematically assess whether the main-belt can dynamically supply objects onto Kamo`oalewa-like orbits.

    Our simulations show that a fraction of particles from the $\nu_6$,
    the 3:1J MMR, and the Flora family can evolve onto Kamo`oalewa-like orbits,
    with transfer efficiencies of 3.31\%, 0.39\%, and 2.54\%, respectively. These findings
    indicate that the main-belt, particularly the low-inclination inner region, can serve as a
    viable source of Kamo`oalewa-like objects. We further examine representative
    dynamical pathways: particles from the Flora family and near the $\nu_6$ are mainly delivered
    via the $\nu_6$, whereas those near the 3:1J MMR follow more diverse routes, through either the 3:1J~MMR or the $\nu_6$. We note that these pathways are illustrative and do not exhaust all possible evolutionary scenarios.

    Overall, our results show that the Moon is not the only dynamically accessible source of
    Kamo`oalewa \citep{Jiao2024, Zhu2025}. We identify representative transport pathways from main-belt candidate regions
    that deliver particles onto Kamo`oalewa-like orbits.
    {It should be noted that NEOs sources are distributed across the entire main-belt
    (e.g., \citealt{Morbidelli2002, Granvik2017, Zhou2024, Zhou2025}).
    However, here we focus on dynamically efficient pathways/regions that are most
    relevant to producing Kamo`oalewa. Within the adopted
    integration timescale of 100~Myr, classical strong resonances such as $\nu_6$ and 3:1J MMR
    represent the most efficient routes for transporting main-belt bodies into the near-Earth
    region. As a first step, we therefore systematically examine a limited
    number of the most probable and efficient pathways relevant to the formation of  Kamo`oalewa-like orbits.
    Future work will address the transition from this orbital domain to the current co-orbital
    state of Kamo`oalewa through more restrictive criteria and dedicated analysis.}

    {
        Over the past decades, a number of asteroid exploration missions have significantly enhanced
        our understanding of small bodies through a combination of rendezvous, sample-return, and
        flyby observations.
        In addition to dedicated missions such as Hayabusa, Hayabusa2 and OSIRIS-REx, which
        explored the near-Earth asteroids like Itokawa, Ryugu, and Bennu
        \citep{Fujiwara2006,Yano2006,Watanabe2017,Lauretta2017,Lauretta2019,Watanabe2019,
        Kitazato2019,Hamilton2019,Aokietal2020,Potiszil2023,Glavinetal2025},
        respectively, several spacecraft also have performed
        close flybys of small bodies, including the Chang'e-2 flyby of Toutatis
        (\citealt{Huang2013,Jiang2015} and references therein).
        In this context, as the first sample-return mission to explore a small, rapidly rotating asteroid, the
        {\emph{Tianwen-2}} mission \citep{Zhang2019,Zhang2025b,YING2025} is expected to
        provide valuable insights into surface environments, interior structure and dynamical properties.
    }

\begin{acknowledgements}
        We sincerely thank the referee for the valuable comments and constructive suggestions that
        have improved the quality of the manuscript.
        This work is financially supported by the National Natural Science Foundation of China
        (grant Nos. 12561160085 and 12150009), and the Foundation of Minor Planets of the Purple Mountain Observatory.
        This research has made use of data provided by the International Astronomical Union's Minor Planet Center.
\end{acknowledgements}

\bibliographystyle{raa}
\bibliography{ms}

\end{document}